\documentclass[12pt,preprint]{aastex}

\newcommand{\psr}{\mbox{PSR J0437$-$4715}}

\newcommand{\pb}{\mbox{$P_{\rm b}$}}
\newcommand{\pbdot}{\mbox{$\dot{\pb}$}}

\shorttitle{High-Precision Timing of \psr}
\shortauthors{Verbiest et al.}

\begin{document}

\title{Precision timing of \psr: an accurate pulsar distance,
  a high pulsar mass and a limit on the variation of Newton's
  gravitational constant}

\author{J. P. W. Verbiest,\altaffilmark{1,2}
  M. Bailes,\altaffilmark{1} W. van Straten,\altaffilmark{1}
  G. B. Hobbs,\altaffilmark{2} R. T. Edwards,\altaffilmark{2}
  R. N. Manchester,\altaffilmark{2} N. D. R. Bhat,\altaffilmark{1}
  J. M. Sarkissian,\altaffilmark{2} B. A. Jacoby\altaffilmark{3} and
  S. R. Kulkarni\altaffilmark{4}} 

\altaffiltext{1}{Centre for Astrophysics and Supercomputing, Swinburne
  Unversity of Technology, P.O. Box 218 Hawthorn, VIC 3122, Australia}
  \altaffiltext{2}{Australia Telescope National Facility -- CSIRO,
  P.O. Box 76, Epping, NSW 1710, Australia} \altaffiltext{3}{Naval
  Research Laboratory, 4555 Overlook Avenue, SW, Washington, DC 20375}
  \altaffiltext{4}{Robinson Laboratory, California Institute of
  Technology, Mail Code 105-24, Caltech Optical Observatories,
  Pasadena, CA 91125}

\begin{abstract}
Analysis of ten years of high-precision timing data on the millisecond
pulsar \psr\ has resulted in a model-independent kinematic distance
based on an apparent orbital period derivative, \pbdot, determined at
the $1.5\%$ level of precision ($D_{\rm k} = 157.0\pm 2.4$\,pc),
making it one of the most accurate stellar distance estimates
published to date.  The discrepancy between this measurement and a
previously published parallax distance estimate is attributed to
errors in the DE200 Solar System ephemerides. The precise measurement
of \pbdot\ allows a limit on the variation of Newton's gravitational
constant, $|\dot{G}/G| \leq 23 \times 10^{-12}$\,yr$^{-1}$.  We also
constrain any anomalous acceleration along the line of sight to the
pulsar to $|a_{\odot}/c| \leq 1.5\times 10^{-18}$\,s$^{-1}$ at $95\%$
confidence, and derive a pulsar mass, $m_{\rm psr} = 1.76 \pm
0.20\,M_{\odot}$, one of the highest estimates so far obtained.
\end{abstract}


\section{Introduction}\label{Introduction}

In \citeyear{jlh+93}, \citeauthor{jlh+93} reported the discovery of
\psr, the nearest and brightest millisecond pulsar known. Within a
year, the white dwarf companion and pulsar wind bow shock were
observed \citep{bbb93} and pulsed X-rays were detected
\citep{bt93a}. The proper motion and an initial estimate of the
parallax were later presented along with evidence for secular change
in the inclination angle of the orbit due to proper motion
\citep{sbm+97}. Using high time resolution instrumentation, the
three-dimensional orbital geometry of the binary system was
determined, enabling a new test of general relativity
\citep[GR;][]{vbb+01}. Most recently, multi-frequency observations
were used to compute the dispersion measure structure function
\citep{yhc+07}, quantifying the turbulent character of the
interstellar medium towards this pulsar.

The high proper motion and proximity of \psr\ led to the prediction
\citep{bb96} that a distance measurement independent of parallax would
be available within a decade, when the orbital period derivative
(\pbdot) would be determined to high accuracy. Even if the predicted
precision of $\sim$1\% would not be achieved, such a measurement would
be significant given the strong dependence of most methods of distance
determination on relatively poorly constrained models and the
typically large errors on parallax measurements. Even for nearby
stars, both the Hubble Space Telescope and the Hipparcos satellite
give typical distance errors of $3\%$ \citep{val07} and so far only
two distances beyond $100$\,pc have been determined at $\sim1\%$
uncertainty \citep{tlmr07}. This kinematic distance is one of the few
model-independent methods that does not rely upon the motion of the
Earth around the Sun.

As demonstrated by \citet{dt91}, \pbdot\ can also be used to constrain
the variation of Newton's gravitational constant. The best such limit
from pulsar timing to date \citep[$|\dot{G}/G| = (4\pm5)\times
10^{-12}$\,yr$^{-1}$ from PSR B1913+16]{tay93} is compromised due to
the poorly constrained equation of state for the neutron star
companion \citep{nor90}. The slightly weaker but more reliable limit
of $|\dot{G}/G| = (-9\pm 18) \times 10^{-12}$\,yr$^{-1}$ \citep[from
PSR B1855+09, which has a white dwarf companion]{ktr94} should
therefore be considered instead. A more stringent limit can be
obtained from the \citet{sns+05} timing of PSR J1713+0747:
$|\dot{G}/G| = (2.5\pm 3.3) \times 10^{-12}$\,$yr^{-1}$ (at 95\%
certainty). This limit is, however, based upon the formal errors of
\pbdot, $P_{\rm b}$ and parallax, which are easily underestimated by
standard methodologies, as we shall demonstrate later. Because of this
we believe the \citet{sns+05} limit is probably underestimated, but
still of relevance.  However, none of these limits are as strong as
that put by lunar laser ranging \citep[LLR;][]{wtb04}: $\dot{G}/G =
(4\pm 9) \times 10^{-13}$\,yr$^{-1}$. Besides limiting alternative
theories of gravity, bounds on $\dot{G}$ can also be used to constrain
variations of the Astronomical Unit ($AU$). Current planetary radar
experiments \citep{kb04} have measured a significant linear increase
of $d AU/dt =0.15\pm 0.04$\,m yr$^{-1}$, which may imply $\dot{G}/G =
(-10 \pm 3)\times 10^{-13}$\,yr$^{-1}$, just beyond the sensitivity of
the limits listed above.

As mentioned before, the equation of state for dense neutron star
matter is very poorly constrained.  Pulsar mass determinations can
probe the range of permissible pulsar masses and thereby limit
possible equations of state \citep{lp07}.  Presently, only the pulsars
NGC 6440B, Terzan 5\,I and Terzan 5\,J have predicted masses higher
than the typical value of $1.4\,M_{\odot}$\citep{frb+07,rhs+05};
however, as discussed in more detail in Section \ref{Mass}, such
predictions do not represent objective mass estimates.

The remainder of this paper is structured as follows: Section
\ref{Obs} describes the observations, data analysis and general timing
solution for \psr. Section \ref{Dist} describes how the measurement of
\pbdot\ leads to a new and highly precise distance. In Section
\ref{GandA}, this measurement is combined with the parallax distance
to derive limits on $\dot{G}$ and the Solar System
acceleration. Section \ref{Mass} presents the newly revised pulsar
mass and our conclusions are summarised in Section \ref{Conc}.

\section{Observations and Data Reduction}\label{Obs}

\begin{figure}
\includegraphics[width=8.2cm]{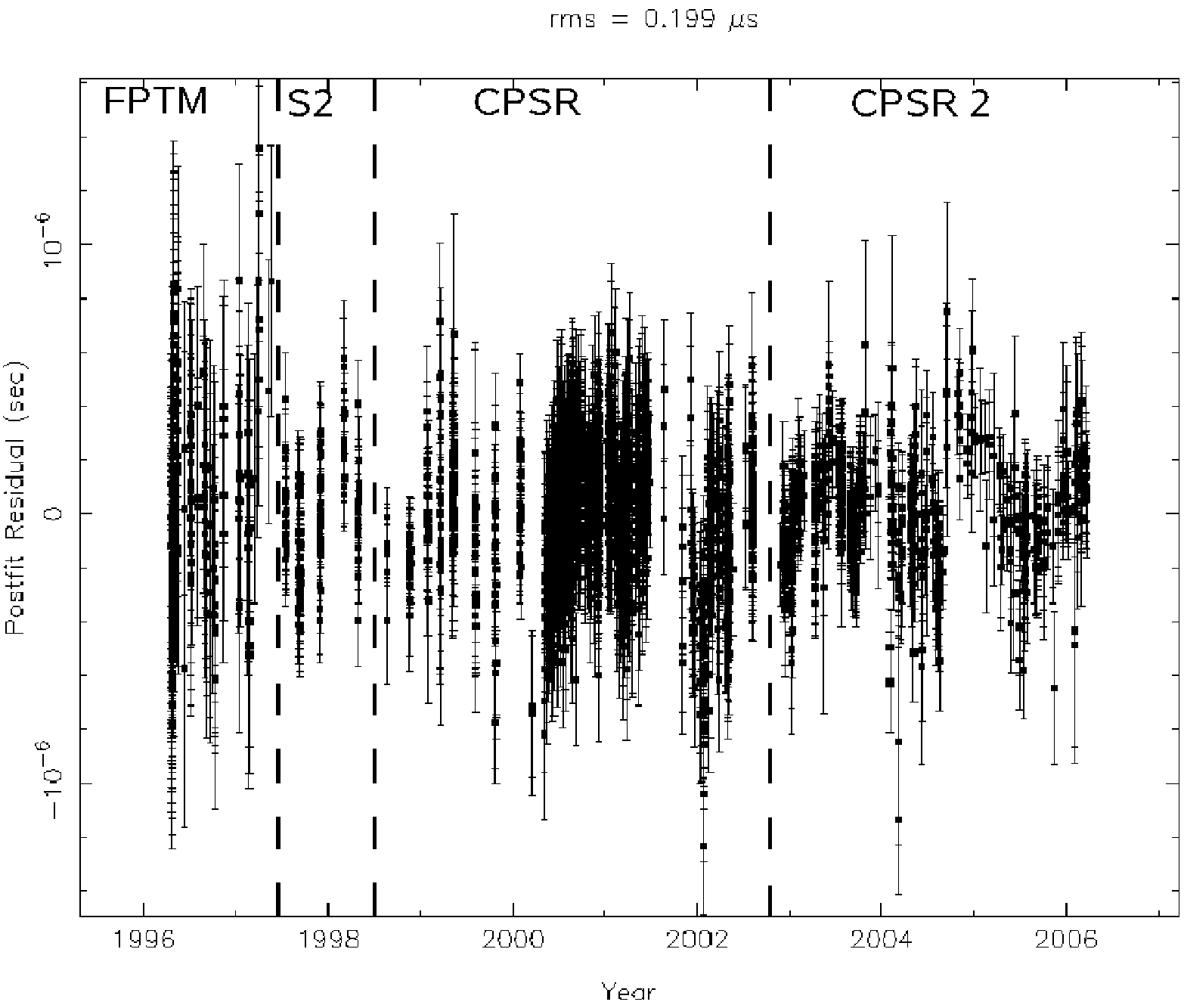}
\caption{Combined 20\,cm post-fit timing residuals for new and
  archival \psr\ timing data. Vertical dashed lines separate the
  different instruments.\label{Fig::Res}}
\end{figure}

Observations of \psr\ were made over a time span of ten years (see
Figure \ref{Fig::Res}), using the Parkes 64-m radio telescope, two
20\,cm receiving systems (the central beam of the Parkes multi-beam
receiver \citep{swb96} and the H-OH receiver) and four generations of
digital instrumentation (see Table \ref{tbl::Obs}): the Fast Pulsar
Timing Machine (FPTM), the S2 VLBI recorder, and the
Caltech-Parkes-Swinburne Recorders (CPSR and CPSR2).  The FPTM is an
autocorrelation spectrometer, whereas the three other instruments are
baseband data recording and processing systems that employ
phase-coherent dispersion removal.

\subsection{Arrival Time Estimation}
For the FPTM, S2, and CPSR backends, the uncalibrated polarization
data were combined to form the polarimetric invariant interval
\citep{bri00} and each observation was integrated in time and
frequency before pulse arrival times were calculated through standard
cross-correlation with an instrument-dependent template profile.  For
the CPSR2 data, the technique described by \citet{van04a} was used to
calibrate 5 days of intensive \psr\ observations made on 2003 July 19
to 21, 2003 August 29, and 2005 July 24.  The calibrated data were
integrated to form a polarimetric template profile with an integration
length of approximately 40 hours and frequency resolution of 500 kHz.
This template profile and Matrix Template Matching
\citep[MTM,][]{van06} were used to calibrate the three years of CPSR2
data.  An independent MTM fit was performed on each five-minute
integration, producing a unique solution in each frequency channel, as
shown in Figure 2 of \citet{van06}. The calibrated data were then
integrated in frequency to produce a single full-polarization profile
at each epoch.  MTM was then used to derive time-of-arrival (TOA)
estimates from each calibrated, five-minute integration.  The
application of MTM during the calibration and timing stages reduced
the weighted rms of the CPSR2 post-fit timing residuals by a factor of
two.  All the data reduction described above was performed using the
\textsc{psrchive} software package \citep{hvm04}.

\subsection{Timing Analysis}
Most data were recorded at a wavelength of 20\,cm; however, in the
final three years, simultaneous observations at 10 and 50\,cm were
used to measure temporal variations of the interstellar dispersion
delay \citep[corrections for these variations were implemented in a
way similar to that of][]{yhc+07}. A linear trend of these delays was
also obtained for the year of FPTM data, using data at slightly
different frequencies close to $1400$\,MHz.

The arrival times were analyzed using the \textsc{Tempo2} pulsar
timing software package \citep{hem06,ehm06} and consistency with the
earlier program, \textsc{Tempo}, was verified.  The timing model (see
Table \ref{Model}) is based on the relativistic binary model first
derived by \citet{dd86}, and expanded to contain the geometric orbital
terms described by \citet{kop95,kop96}. The model is optimised through
a standard weighted least-squares fit in which all parameters are
allowed to vary, including the unknown time delays between data from
different instruments, but excluding the mean value of dispersion
measure, which is determined from the simultaneous CPSR2, 64\,MHz-wide
bands centred at 1341 and 1405\,MHz.

A major difference between our implementation of solutions for the
orbital angles $\Omega$ and $i$ and previous efforts
\citep{vbb+01,hbo06} is that they were implemented as part of the
standard fitting routine.  This ensures any covariances between these
and other parameters (most importantly the periastron advance and
companion mass, see Table \ref{Model} and Section \ref{Mass}) are
properly accounted for, thereby yielding a more reliable measurement
error. The previous works 
mentioned above derived these effects from an independent mapping of
$\chi^2$ space, leaving the errors of other parameters unaffected.

\begin{figure}
\includegraphics[width=6.15cm,angle=-90.0]{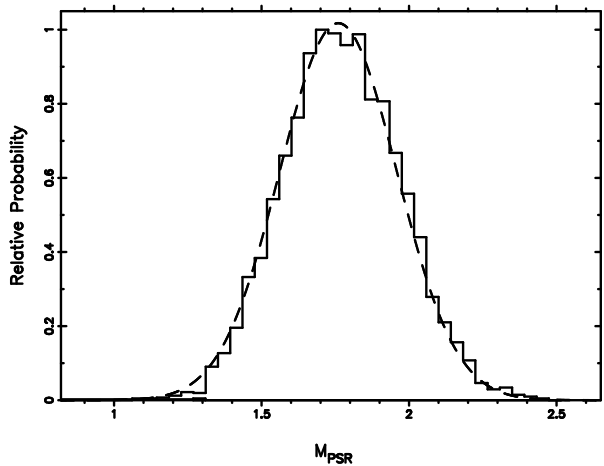}
\caption{Pulsar mass probability distribution. The solid line shows the
  histogram of 5000 pulsar masses derived from a Monte-Carlo
  simulation with power spectrum and sampling equal to that of the
  \psr\ data set. The dashed line is a Gaussian distribution with a
  mean value of $m_{\rm psr} = 1.76\,M_{\odot}$ and standard deviation
  of $0.20\,M_{\odot}$. 
  \label{PSRMass}}
\end{figure}

As can be seen from Figure \ref{Fig::Res}, there are significant
low-frequency structures present in the timing residual data. Since
the standard least-squares fitting routine used in \textsc{Tempo2}
does not account for the effect of such correlations on parameter
estimation, we performed a Monte-Carlo simulation where data sets with
a post-fit power spectrum statistically consistent with that of the
\psr\ data were used to determine the parameter estimations
uncertainties in the presence of realistic low frequency noise. These
errors, as well as the factors by which the original errors were
underestimated, are shown in Table \ref{Model}. As an example, the
distribution of derived pulsar masses from the Monte-Carlo simulation
is given in Figure \ref{PSRMass}. Because of the dispersion measure
corrections implemented in the final three years of data, one can
expect the spectrum of these most precise data points to contain less
low-frequency noise than the ten year data set as a whole. We
therefore expect the errors resulting from this analysis to be
slightly overestimated. Ongoing research into extending the fitting
routine with reliable whitening
schemes to avoid the spectral leakage and hence improve the
reliability of the measured parameters, is expected to reduce these
errors by factors of around two. All errors given in this paper are
those resulting from the Monte-Carlo simulations, unless otherwise
stated. The simulations also showed that any biases resulting from the
red noise are statistically negligible for the reported parameters. (A
full description of this Monte-Carlo technique and the whitening
schemes mentioned will be detailed in a future publication.)

\subsection{Solar System Ephemerides}\label{SSE}

Pulsar timing results are dependent on accurate ephemerides for the
Solar System bodies. The results presented in this paper were obtained
using the DE405 model \citep{sta04b} and, for comparison, selected
parameters obtained with the earlier DE200 model are shown in Table
\ref{DEModels}. The greatly reduced $\chi^2$ indicates that the newer
Solar System ephemerides are superior to the earlier DE200,
reinforcing similar conclusions of other authors
\citep{sns+05,hbo06}. We notice the parallax value changes by more
than $10\,\sigma$, and that the different derived values are closely
correlated with the ephemeris used. Although the effect is not as
dramatic as it appears because of the under-estimation of the
\textsc{Tempo2} errors, the fact that the DE405 results agree much
better with the more accurate kinematic distance (discussed in the
next Section), strongly suggests that the differences are due to the
ephemeris used and confirms that the DE405 ephemeris is superior.
Finally, we note that the DE405 measurement of $\dot{\omega}$ ($0.016
\pm 0.008\,^{\circ}{\rm yr}^{-1}$) is consistent with the GR
prediction for this system ($0.0172 \pm 0.0009\,^{\circ}{\rm
yr}^{-1}$).

\section{Kinematic Distance}\label{Dist}

\begin{figure}
\includegraphics[width=8.2cm]{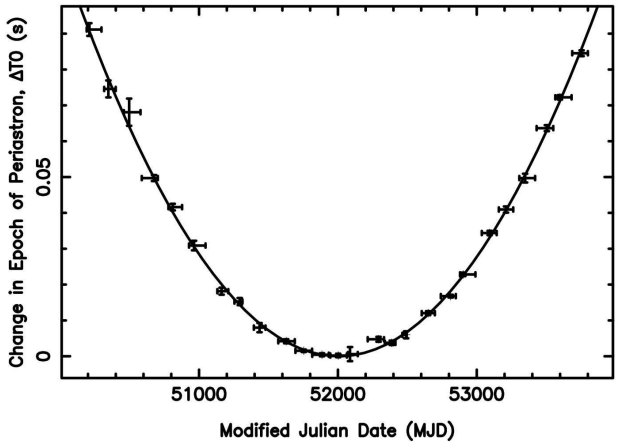}
\caption{Variations in epoch of periastron passage ($T_{\rm 0}$) due
  to apparent orbital period increase. A steady increase in orbital
  period is equivalent to a quadratic increase in $T_{\rm 0}$ relative
  to periastron times for a constant orbital period.  For this plot,
  $T_{\rm 0}$ was measured on data spans of up to 120 days with a
  model having no orbital period derivative. The formal one-$\sigma$
  measurement errors reported by \textsc{Tempo2} are shown by vertical
  error bars and the epochs 
  over which the measurements were made are shown by horizontal bars. As
  the mean measurement time was determined through a weighted average
  of the data contained in the fit, these horizontal bars need not be
  centred at the mid time associated with the measurement.  
  The parabola shows the effect of the \pbdot\
  value obtained from a fit to the data shown in Figure
  \ref{Fig::Res}.\label{Fig::Pbdot}}
\end{figure}

As shown in Figure \ref{Fig::Pbdot}, the long-term timing history
enables precise measurement of the orbital period derivative, \pbdot $
= (3.73\pm0.06)\times 10^{-12}$.  This observed value represents a
combination of phenomena that are intrinsic to the binary system and
dynamical effects that result in both real and apparent accelerations
of the binary system along the line of sight \citep{bb96}; i.e.
\begin{equation}\label{PbdotEq::Basis}
\dot{P}_{\rm b}^{\rm obs} =
\dot{P}_{\rm b}^{\rm int} +
\dot{P}_{\rm b}^{\rm Gal} +
\dot{P}_{\rm b}^{\rm kin}
\end{equation}
where ``obs'' and ``int'' refer to the observed and intrinsic values;
``Gal'' and ``kin'' are the Galactic and kinematic contributions.

Intrinsic orbital decay is a result of energy loss typically due to
effects such as atmospheric drag and tidal dissipation; however, in a
neutron star--white dwarf binary system like \psr, energy loss is
dominated by quadru\-po\-lar gravitational wave emission. For this
system, GR predicts \citep{tw82} $\dot P_{\rm b}^{\rm
GR}=-4.2\times10^{-16}$, two orders of magnitude smaller than the
uncertainty in the measured value of \pbdot.

Galactic contributions to the observed orbital period derivative
include differential rotation and gravitational acceleration
\citep{dt91}.  The differential rotation in the plane of the Galaxy is
estimated from the Galactic longitude of the pulsar and the
Galactocentric distance and circular velocity of the Sun. Acceleration
in the Galactic gravitational potential varies as a function of height
above the Galactic plane \citep{hf04b}, which may be estimated using
the parallax distance and the Galactic latitude of the
pulsar. Combining these terms gives $\dot{P}_{\rm b}^{\rm Gal} = (-1.8
-0.5) \times 10^{-14} = -2.3\times10^{-14}$, which is of the same
order as the current measurement error.

\begin{figure}
\includegraphics[width=8.2cm]{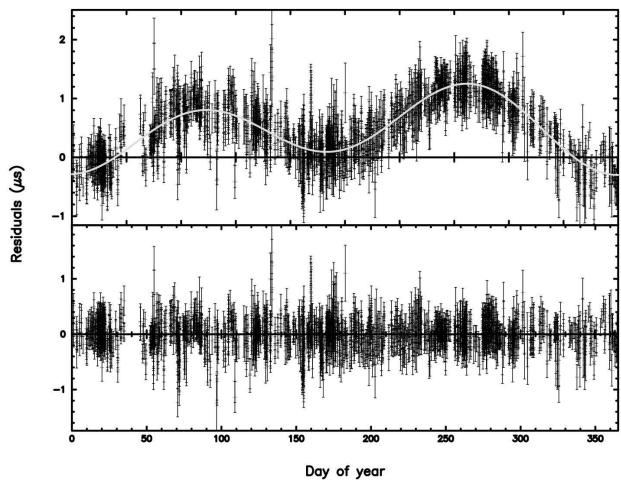}
\caption{Parallax signature of \psr. Top: Timing residuals for \psr\
  as a function of day of year (starting on 18 November), without
  parallax but with all remaining parameters at their best-fit
  values. The smooth curve represents the model fit of a parallax of
  6.65 mas. Bottom: The same timing residuals with parallax included
  in the model. The overall rms for the top and bottom plots is $524$
  and $199$\,ns respectively. 
  The double-humped signature specific
  to parallax originates from the delay in pulse time-of-arrival (TOA)
  as 
  the Earth orbits the Sun and samples different parts of the curved
  wave-front originating at the pulsar.\label{Fig::Px}}
\end{figure}

Given the negligible intrinsic contribution, Equation
\ref{PbdotEq::Basis} can be simplified and rewritten in terms of the
dominant kinematic contribution known as the Shklovskii effect
\citep{shk70}, an apparent acceleration resulting from the non-linear
increase in radial distance as the pulsar moves across the plane
perpendicular to the line of sight; quantified by the proper motion,
$\mu$, and distance $D$ from the Earth:
\begin{equation}\label{PbdotEq::Shk}
\dot{P}_{\rm b}^{\rm obs} -
\dot{P}_{\rm b}^{\rm Gal} \simeq
\dot{P}_{\rm b}^{\rm kin} = \frac{\mu^2D}{c} P_{\rm b},
\end{equation}
where $c$ is the vacuum speed of light. Using the measured values of
$\mu$, \pb, and \pbdot, Equation \ref{PbdotEq::Shk} is used to derive
the kinematic distance \citep{bb96}: $D_k = 157.0 \pm 2.4$ pc.  This
distance is consistent with the one derived from parallax ($D_{\rm \pi}
= 150\pm12
$\, pc -- see also Figure \ref{Fig::Px}) and is, with a relative error
of $1.5\%$, comparable in precision to the best parallax measurements
from VLBI \citep{tlmr07} and better than typical relative errors
provided by the Hipparcos and Hubble space telescopes \citep{val07}.

Given the dependence of parallax distances on ephemerides, as
described in Section \ref{SSE}, it is interesting to note the
robustness of $D_{\rm k}$.  Also, Table \ref{Model} shows that the
presence of red noise corrupts the parallax error by a factor of 7.9,
whereas \pbdot\ is only affected by a factor of 2.5.  These facts
clearly indicate the higher reliability of $D_{\rm k}$ as compared to
$D_{\rm \pi}$.

\section{Limits on \pbdot\ Anomalies: $\dot{G}$ and the
  Acceleration of the Solar System}\label{GandA}

Any anomalous orbital period derivative can be constrained by
substituting the parallax distance into Equation 2, yielding
\begin{eqnarray}
  \Big(\frac{\dot{P}_{\rm b}}{P_{\rm b}}\Big)^{\rm excess} & = &
  \big(\dot{P}_{\rm b}^{\rm obs} - \dot{P}_{\rm b}^{\rm Gal} -
  \dot{P}_{\rm b}^{\rm kin}\big)/P_{\rm b}\nonumber\\ 
  & = & (3.2 \pm 5.7) \times 10^{-19}\, {\rm s}^{-1}.
\end{eqnarray}
in which the error is almost exclusively due to the parallax
uncertainty. Following \citet{dt91}, this can be translated into a
limit on the time derivative of Newton's gravitational constant (given
are $95\%$ confidence levels):
\begin{equation}
\frac{\dot{G}}{G} = -\frac{1}{2}\Big(\frac{\dot{P_{\rm b}}}{P_{\rm
    b}}\Big)^{\rm excess} = (-5\pm 18) \times 10^{-12}\, {\rm yr}^{-1}
\end{equation}

This limit is of the same order as those previously derived from
pulsar timing (see Section \ref{Introduction}), but a currently
ongoing VLBI campaign on this pulsar is expected to improve
significantly on our parallax measurement, and this should improve our
limit to close to that put by LLR \citep[\mbox{($4\pm 9) \times
10^{-13}$\, yr$^{-1}$};][]{wtb04}.  The LLR experiment is based on a
complex $n$--body relativistic model of the planets that incorporates
over 140 estimated parameters, such as elastic deformation, rotational
dissipation and two tidal dissipation parameters.  In contrast, the
\psr\ timing result is dependent on a different set of models and
assumptions, and therefore provides a useful independent confirmation
of the LLR result.

A recent investigation into the possible causes of a measured
variability of the Astronomical Unit \citep[$AU$;][]{kb04} has refuted
all but two sources of the measured value of $d AU/dt = 0.15 \pm
0.04\, {\rm m/yr}$. \citet{kb04} state that the measured linear
increase in the $AU$ would be due to either systematic effects or to a
time-variation of $G$ at the level of $\dot{G}/G = (-10\pm 3) \times
10^{-13}\, {\rm yr}^{-1}$, comparable to, but inconsistent with, the
LLR limit.

The anomalous \pbdot\ measurements of a number of millisecond pulsars
have also been used to place limits on the acceleration of the Solar
System due to any nearby stars or undetected massive planets
\citep{zt05}. The \psr\ data set limits any anomalous Solar System
acceleration to $ | a_{\odot}/c | \leq 1.5\times 10^{-18}\, {\rm
s}^{-1}$ in the direction of the pulsar with $95$\% certainty.  This
rules out any Jupiter-mass planets at distances less than $117\,$AU
along the line of sight, corresponding to orbital periods of up to
1270\,years. Similarly, this analysis excludes any Jupiter-mass
planets orbiting \psr\ between $\sim$5 and $117\,$AU along the line of
sight. \citet{zt05} also compared the sensitivity of this limit to
that of optical and infra-red searches for trans-Neptunian objects (TNOs) and
concluded that beyond $\sim 300$\,AU the acceleration limit becomes
more sensitive than the alternative searches. At a distance of
$300\,$AU from the Sun, the $95\%$ confidence upper limit on the mass
of a possible TNO (in the direction of the pulsar) is $6.8$ Jupiter
masses. The precise VLBI measurement of parallax mentioned above
might decrease this to close to one Jupiter mass.

\section{Pulsar Mass}\label{Mass}

A combination of the mass function and a measurement of the Shapiro
delay range can be used to obtain a measurement of the pulsar
mass. Using this method, \citet{vbb+01} derived a mass for \psr\ of
$1.58 \pm 0.18\,M_{\odot}$ whereas \citet{hbo06} obtained $1.3 \pm
0.2\,M_{\odot}$. It should be noted, however, that these values
resulted from a model that incorporated geometric parameters first
described by \citet{kop95,kop96}, but covariances between these and
other timing parameters (most importantly the companion mass or
Shapiro delay range) were not taken into account. Whilst the length of
the data sets used by these authors were only a few years, it can also
be expected that some spectral leakage from low-frequency noise was
unaccounted in the errors of these previously published values. As
described in Section \ref{Obs}, the Monte-Carlo simulations and
extended fitting routines implemented for the results reported in this
paper do include these covariances and spectral leakage; it can
therefore be claimed that the current estimates (at $68\%$ confidence)
of $m_{\rm c} = 0.254 \pm 0.018\,M_{\odot}$ and $m_{\rm psr} = 1.76
\pm 0.20\,M_{\odot}$, for the white dwarf companion and pulsar
respectively, reflect the measurement uncertainty more realistically
than any previous estimate.  The distribution of $m_{\rm psr}$ that
follows from the 5000 Monte-Carlo realizations is shown in Figure
\ref{PSRMass}, together with a Gaussian with mean $1.76$ and standard
deviation $0.20$. This demonstrates the symmetric distribution of the
pulsar mass likelihood distribution, induced by the precise
determination of the orbital inclination angle.

We also note that the new mass measurement of \psr\ is the highest
obtained for any pulsar to date. Distinction needs to be made between
the objective mass estimate presented in this paper and the subjective
mass predictions presented in \citet{rhs+05} and \citet{frb+07}. The
pulsar mass confidence interval presented in this paper is derived
from the measurement uncertainties of all relevant model parameters,
including the well-determined orbital inclination angle, $i$. In
contrast, $i$ is unknown in the Terzan 5\,I and J \citep{rhs+05} and
PSR J1748$-$2021B \citep{frb+07} binary systems, and the posterior
probability intervals for the pulsar masses presented in these works
are based upon the prior assumption of a uniform distribution of $\cos
i$. These fundamental differences must be accounted for in any
subsequent hypothesis testing. 
Consequently, \psr\ is currently the only pulsar to provide reliable
constraints on equations of state based on hyperons and Bose-Einstein
condensates as described by \citet{lp07}. Simulations with
\textsc{Tempo2} indicate that a forthcoming observational campaign
with a new generation of backend systems can be expected to increase
the significance of this measurement by another factor of about two in
the next year.
 
\section{Conclusions}\label{Conc}

We have presented results from the highest-precision long-term timing
campaign to date. With an overall residual rms of 199\,ns, the
10\,years of timing data on \psr\ have provided a precise measurement
of the orbital period derivative, \pbdot, leading to the first
accurate kinematic distance to a millisecond pulsar: $D_{\rm k} =
157.0\pm 2.4$\,pc. Application of this method to other pulsars in the
future can be expected to improve distance estimates to other binary
pulsar systems \citep{bb96}.

Another analysis based on the \pbdot\ measurement places a limit on
the temporal variation of Newton's gravitational constant. We find a
bound comparable to the best so far derived from pulsar timing:
$\dot{G}/G = (-5\pm18)\times 10^{-12}\,{\rm yr}^{-1}$.  An ongoing
VLBI campaign on this pulsar is expected to improve this limit,
enabling an independent confirmation of the LLR limit.

Previous estimates of the mass of \psr\ have been revised upwards to
$m_{\rm psr} = 1.76 \pm 0.20\, M_{\odot}$, which now makes it one of
the few pulsars with such a heavy mass measurement.  A new generation
of backend instruments, dedicated observing campaigns, and data
prewhitening techniques currently under development should decrease
the error in this measurement enough to significantly rule out various
equations of state for dense nuclear matter.

\acknowledgments

The Parkes Observatory is part of the Australia Telescope which is
funded by the Commonwealth of Australia for operation as a National
Facility managed by CSIRO.  We thank the staff at Parkes Observatory
for technical assistance during regular observations. B.A.J holds a
National Research Council Research Associateship Award at the Naval
Research Laboratory (NRL). Basic research in radio astronomy at NRL is
supported by the Office of Naval Research. The authors wish to express
their gratitude to William Coles from the University of California at
San Diego, for extensive discussion and help with the Monte-Carlo
error analysis. Finally, we thank the referee for valuable and
inspiring comments.

{\it Facilities:} \facility{Parkes}

\clearpage

 \clearpage
 
\begin{deluxetable}{cclrrccl}
\tablecaption{Characteristics of the timing data for the four
  instruments used\label{tbl::Obs}}
\tablewidth{0pt}
\tablehead{
\colhead{Backend} & \colhead{Date range} & \colhead{Ref.} &
\colhead{Bandwidth} & \colhead{RMS} & \colhead{Observation} &
\colhead{Number of} & \colhead{TOA} \\
\colhead{} & \colhead{} & \colhead{} &
\colhead{} & \colhead{Residual} & \colhead{length\tablenotemark{a}} &
\colhead{TOAs} & \colhead{error\tablenotemark{a}} 

}
\tablecolumns{8}
\startdata
FPTM & 1996 Apr -- 1997 May & 1,2 & $256\,$MHz &
  $368\,$ns & $10\,$min & 207 & $500\,$ns \\
S2 & 1997 Jul -- 1998 Apr & 3 & $16\,$MHz & $210\,$ns
  & $120\,$min & 117 & $160\,$ns \\
CPSR & 1998 Aug -- 2002 Aug & 3 & $20\,$MHz &
  $218\,$ns & $15\,$min & 1782 & $250\,$ns \\
CPSR2 & 2002 Nov -- 2006 Mar & 4 & $2\times
64\,$MHz\tablenotemark{b} &
  $164\,$ns & $60\,$min & 741 & $150\,$ns \\
\enddata
\tablenotetext{a}{Displayed are typical values only.}
\tablenotetext{b}{CPSR2 records two adjacent $64\,$MHz bands
  simultaneously at 20\,cm.}
\tablerefs{
  (1) \citet{sbm+97}; (2) \citet{san01}; (3) \citet{van03a}; (4)
  \citet{hbo06}.}
\end{deluxetable}

\clearpage

\begin{deluxetable}{llrcr}
\tablecaption{\psr\ timing model
  parameters\tablenotemark{a}\label{Model}} 
\tablewidth{0pt}
\tablehead{
  \colhead{Parameter Name} & \colhead{Parameter} &
  \colhead{\textsc{Tempo2}}& \colhead{Monte-Carlo} & \colhead{Error}\\ 
  \colhead{and Units} & \colhead{Value} &
  \colhead{Error\tablenotemark{b}} & \colhead{Error\tablenotemark{b}}
  & \colhead{Ratio} 
}
\tablecolumns{2}
\startdata
\cutinhead{Fit and Data Set}
MJD range \dotfill & 50191.0--53819.2 & & & \\
Number of TOAs \dotfill & 2847 & & & \\
Rms timing residual ($\mu$s) \dotfill & 0.199 & & & \\
\cutinhead{Measured Quantities}
  Right ascension, $\alpha$ (J2000)       \dotfill &
  04$^{\mathrm h}$37$^{\mathrm m}$15\fs8147635 & 3 & 29 & 9.8\\
  Declination, $\delta$ (J2000)           \dotfill & 
  $-$47\degr15\arcmin08\farcs624170 & 3 & 34 & 11\\
  Proper motion in $\alpha$, $\mu_\alpha \cos{\delta}$ (mas yr$^{-1}$)
  \dotfill &  121.453 & 1 & 10 & 8.7 \\ 
  Proper motion in $\delta$, $\mu_\delta$ (mas yr$^{-1}$) \dotfill &
  $-$71.457 & 1 & 12 & 9.0\\ 
  Annual parallax, $\pi$ (mas)  \dotfill & 6.65 & 7 & 51 & 7.9 \\
  Dispersion measure, $DM$ (cm$^{-3}$ pc) \dotfill & 2.64476 & 7 &
  \tablenotemark{d} & \tablenotemark{d} \\
  Pulse period, $P$ (ms)                    \dotfill &
  5.757451924362137 & 2 & 99 & 47\\ 
  Pulse period derivative, $\dot{P}$ (10$^{-20}$) \dotfill &
  5.729370 & 2 & 9 & 4.8 \\
  Orbital period, $P_{\rm b}$ (days)        \dotfill &
  5.74104646\tablenotemark{c} & 108 & 200 & 1.9 \\
  Orbital period derivative, $\dot{P_{\rm b}}$ (10$^{-12}$) \dotfill &
  3.73 & 2 & 6 & 2.5\\
  Epoch of periastron passage, $T_{0}$ (MJD) \dotfill &
  52009.852429\tablenotemark{c} & 582 & 780 & 1.3\\
  Projected semi-major axis, $x$ (s)      \dotfill &
  3.36669708\tablenotemark{c} & 11 & 14 & 1.4\\
  Longitude of periastron, $\omega_0$ (\degr) \dotfill &
  1.2224\tablenotemark{c} & 365 & 490 & 1.3 \\
  Orbital eccentricity, $e$ (10$^{-5}$) \dotfill & 1.9180 & 3 &
  7 & 2.1\\ 
  Periastron advance, $\dot \omega$ (\degr\ yr$^{-1}$) \dotfill &
  0.01600\tablenotemark{c} & 430 & 800 & 1.8\\ 
  Companion mass, $m_{\rm 2}$ (M$_{\odot})$       \dotfill &
  0.254\tablenotemark{c} & 14 & 18 & 1.3 \\
  Longitude of ascension, $\Omega$ (\degr)  \dotfill &
  207.8\tablenotemark{c} & 23 & 69 & 3.0 \\
  Orbital inclination, $i$ (\degr)          \dotfill & 137.58 & 6
  & 21 & 3.7\\
\cutinhead{Set Quantities}
Reference epoch for $P$, $\alpha$ & & & & \\
  and $\delta$ determination (MJD)
  \dotfill & 52005 & & & \\
  Reference epoch for DM &  &  & & \\
  determination (MJD) \dotfill & 53211 & & & \\
\enddata
\tablenotetext{a}{These parameters are determined using
  \textsc{Tempo2} which uses the International Celestial Reference
  System and Barycentric Coordinate Time. As a result this timing
  model must be modified before being used with an observing system
  that inputs \textsc{Tempo} format parameters. See \citet{hem06} for
  more information.}
\tablenotetext{b}{Given uncertainties are $1\sigma$ values in the last
  digits of the parameter values.}
\tablenotetext{c}{Because of large covariances, extra precision is given
  for selected parameters.}
\tablenotetext{d}{Dispersion measure was determined through alignment
  of simultaneous CPSR2 observations centred at 1341\,MHz and
  1405\,MHz. The effect of red noise is therefore not applicable.} 
\end{deluxetable}

\clearpage

\begin{deluxetable}{lll}
\tablewidth{0pt} 
\tablecaption{Comparison of DE200 and DE405 results for
  \psr\tablenotemark{a}
  \label{DEModels}}
\tablehead{
\colhead{Parameter name} & \colhead{DE200 result} & \colhead{DE405
  result}}
\startdata
Rms residual (ns) \dotfill & 281 & 199 \\
Relative $\chi^2$ \dotfill & 2.01  & 1.0 \\
Parallax, $\pi$ (mas) \dotfill & 7.84(7)  & 6.65 (7) \\
Parallax distance, $D_{\pi}$ (pc) \dotfill & 127.6(11) & 150.4(16) \\
Previously published $\pi$ (mas) \dotfill 
& 7.19(14)\tablenotemark{1} 
 & 6.3(2)\tablenotemark{2} \\
Kinematic distance, $D_k$ (pc) \dotfill & 154.5 (10) &
156.0 (10) \\  
$D_k$ corrected for Galactic effects (pc) \dotfill & 
   155.5 (10) & 157.0 (10) \\
Variation of Newton's gravitational & & \\
constant, $| {\dot{G} / G} |$ (10$^{-12}$ yr$^{-1}$)\dotfill &
$-21.2(22)$\tablenotemark{b}  & $-5.0(26)$\tablenotemark{b} \\
Total proper motion, $\mu_{tot}$ (mas yr$^{-1}$) \dotfill & 140.852(1)
& 140.915(1) \\ 
Companion mass, $m_{\rm c}$ ($M_{\odot}$) \dotfill & 0.263(14)
 & 0.254(14) \\ 
Pulsar mass, $m_{\rm psr}$ ($M_{\odot}$) \dotfill & 1.85(15)
 & 1.76(15) \\ 
Periastron advance, $\dot{\omega}$ (\degr\ yr$^{-1}$) \dotfill &
0.020(4)  & 0.016(4) \\ 
GR prediction of $\dot{\omega}$ (\degr\ yr$^{-1}$)
\dotfill & 0.0178(9)  & 0.0172(9) 
\enddata
\tablenotetext{a}{Numbers in parentheses represent the formal \textsc{Tempo2}
  $1\,\sigma$ uncertainty in the last digits quoted, unless otherwise stated.}
\tablenotetext{b}{Given are $2\,\sigma$ errors, i.e. $95\%$ confidence levels.}
\tablerefs{
(1) \citet{vbb+01}; (2) \citet{hbo06}}
\end{deluxetable}
\end{document}